\documentclass[twocolumn,prb,10pt,aps,longbibliography,superscriptaddress]{revtex4-2}

\usepackage{graphicx}
\usepackage{amsmath}
\usepackage{amssymb}
\usepackage{amsfonts}
\usepackage{dsfont}
\usepackage{dcolumn}
\usepackage{bbm}
\usepackage{bm}

\usepackage{mathtools}
\usepackage{braket}
\usepackage{physics}
\usepackage{siunitx}
\usepackage[unicode]{hyperref}
\usepackage[nameinlink]{cleveref}
\Crefname{section}{Sec.}{Secs.}
\Crefname{equation}{Eq.}{Eqs.}
\Crefname{figure}{Fig.}{Figs.}
\Crefname{tabular}{Tab.}{Tabs.}

\usepackage{bbold} % For \mathbb{1}
\usepackage{nicefrac}
\usepackage{tikz}
\usetikzlibrary{arrows.meta}
\usepackage{lipsum}  
\usepackage{xcolor}

\usepackage[makeroom]{cancel}

\newcommand{\bs}{\begin{subequations}}
\newcommand{\es}{\end{subequations}}
\newcommand{\be}{\begin{equation}}
\newcommand{\ee}{\end{equation}}

\newcommand{\thr}{\mathrm{thresh}}
\newcommand{\sumnn}{\sum_{\langle i,j\rangle}}

\newcommand{\lr}[1]{\left( #1 \right)}

\newcommand{\lrg}[1]{\left\{#1\right\}}
\newcommand{\lara}[1]{\langle #1 \rangle}

\newcommand{\nbi}[1]{#1_i^\dagger {#1}_i}
\newcommand{\nbk}[1]{#1_\mathbf{k}^\dagger {#1}_\mathbf{k}}

\newcommand{\sumk}{\sum_{\mathbf{k}}}

\newcommand{\ak}{a_\mathbf{k}}
\newcommand{\bk}{b_\mathbf{k}}
\newcommand{\akm}{a_\mathbf{-k}}
\newcommand{\bkm}{b_\mathbf{-k}}

\newcommand{\gam}{\gamma_\mathbf{k}}

\begin{document}
    %==============================================================================
    % Title
    %==============================================================================
    \title{Switching of Magnetization in Quantum Antiferromagnets with Time-Dependent Control Fields }

    %==============================================================================
    % Authors and Affiliation
    %==============================================================================

    \author{Asliddin Khudoyberdiev}
    \email{asliddin.khudoyberdiev@tu-dortmund.de}
    \affiliation{Condensed Matter Theory, 
    TU Dortmund University, Otto-Hahn-Stra\ss{}e 4, 44221 Dortmund, Germany}

    \author{G\"otz S.\ Uhrig}
    \email{goetz.uhrig@tu-dortmund.de}
    \affiliation{Condensed Matter Theory, 
    TU Dortmund University, Otto-Hahn-Stra\ss{}e 4, 44221 Dortmund, Germany}

    \date{\textrm{\today}}

    %==============================================================================
    % Abstract
    %==============================================================================
  \begin{abstract}
Ultrafast manipulation of magnetic states is one of the necessities in modern data storage technology. Quantum antiferromagnets are promising candidates in this respect. The orientation of the order parameter, the sublattice magnetization, can be used to encode ``0'' and ``1'' of a bit. Then, the switching of magnetization and the full control of its reorientation are crucial for writing data. 
We show that the magnetization can be switched efficiently by an 
external short THz pulses with relatively low amplitude. The coupling to
the spin degrees can be direct via the magnetic field or indirect via the electric field
inducing spin-polarized charge currents. 
Our description is based on time-dependent Schwinger boson mean-field theory which includes
the intrinsic dephasing mechanisms beyond a macrospin description. 
The findings help to introduce the use of antiferromagnets in data storage technology.
\end{abstract}

    %Title
    \maketitle

    %==============================================================================
    % Introduction  
    %==============================================================================
    \section{Introduction}
    \label{s:introduction}

Miniaturization and acceleration are two key goals in modern data technology. 
Since long, ferromagnetism has been exploited for data storage and manipulation \cite{chapp07}. 
Recently, antiferromagnetism has attracted much interest for two reasons essentially 
\cite{marro16,jungw16,gomon17}.
First, the absence of stray fields for dense packing of the magnetic domains \cite{loth12}
representing single bits because they do not perturb one another beyond the length
scale set by the domain wall thickness. Second, their generic energy scales are
larger by about a factor 1000 so that ultrafast manipulation comes into reach \cite{gomon14,kampf11}.

If one wants to use the orientation of the sublattice magnetization to encode information
it is crucial to be able to control it. On the one hand, completely spin isotropic systems allow for
ideal, effortless rotation of the orientation. On the other hand, however, they 
possess no robustness against any perturbation, for instance due to noise fields, so that
the information encoded in the orientation is likely to be lost. Thus, one
focuses on systems with a certain degree of spin anisotropy. This, in turn, sets a 
certain activation energy which needs to be overcome to rotate the magnetization.
How this can be done is topic of a plethora of current experimental and theoretical
studies, see Ref.\  \onlinecite{song18} for a review. 

In the present theoretical study, we consider the simple model of an easy-axis anisotropic
spin-$1/2$ Heisenberg model on the square or cubic lattice, i.e., in two or three dimensions.
For simplicity, we incorporate the external control field as an effective, uniform magnetic field
coupled to the localized spins via a Zeeman term. It can be thought of being the magnetic
field component of a THz pulse  as in Refs.\ 
\cite{kampf11,mukai16,jin13} or being generated by the electric field 
inducing currents which in turn exert torques on the spins, spin-transfer torques \cite{katin00}
or spin-orbit torques \cite{liu12a,zelez14,wadle16}. We stress that the time-dependence
of the control field is crucial because it will allow us to adjust the pulses to 
be resonant with the spin anisotropy gap. Being on resonance is a key
aspect in order to manipulate quantum states efficiently and to overcome
activation energies \cite{kampf13,miyas23}.

%techniques

Describing the time evolution of the reorientation of magnetization reliably is far from an easy task.
Brute force numerics quickly comes to its limits because only about 20 to 30 quantum spins can be 
treated which is not sufficient for the treatment of long-range order.
In many studies, the sublattice magnetization is described by a macrospin so that 
antiferromagnetism requires two macrospins, see, e.g., Refs.\ 
\cite{barya80,gomon08,gomon10,gomon14,kim14,kim17}. 
Any dephasing and/or relaxation is incorporated in the standard form of 
Landau-Lifshitz-Gilbert (LLG)
equation with a phenomenological Gilbert damping $\alpha$. Then, one can 
deal with the precessional 
time evolution described by six variables (twice three components) or 
suitable further approximate reductions.

An extension consists in the use of LLG equations for each spin \cite{nowak07,weiss23}. Furthermore,
finite temperature can be accounted for by the Landau-Lifshitz-Bloch equation 
\cite{nowak07,kazan08} which comprises
a stochastic Langevin term to account for thermal fluctuations. Again, these are essentially
classical equations, but with a number of variables scaling linearly with system size.
Indeed, classical spin variable are well suited to capture the dynamics of quantum spins
as well, at least in disordered states \cite{stane14b} which can be justified
by the truncated Wigner approximation \cite{polko10,david15}.
Here, however, we focus on the ordered state mostly at temperature zero.

Since the candidate antiferromagnets for data storage are used deep in their
long-range ordered phases, the application of classical equations of motion
can be questioned. Quantum fluctuations also play a role. The usual way
to capture the leading quantum and low-temperature thermal fluctuations
is spin wave theory about the long-range ordered ground state. The two most
common bosonic representations of spins are the Holstein-Primakoff representation \cite{holst40}
and the Dyson-Maleev representation \cite{dyson56a,malee58} which both describe small
deviation around a specific ordered state. Thus, they are not suitable
to describe large deviations as they necessarily occur in switching processes
where the orientation of the magnetization is rotated by 90$^\circ$ or 
180$^\circ$ so that the state after the switching is as far away from the
initial state as possible. Here a third representation comes in handy which 
is due to Schwinger \cite{schwi52}
allowing to describe disordered spin states or orders with arbitrary orientation
\cite{auerb88,auerb94}.
Very recently, it was shown that a time-dependent version of Schwinger boson 
mean-field theory can capture switching processes \cite{bolsm23}. This approach has the
advantage that the leading quantum and thermal fluctuations are captured.
Not only the modes at zero momentum appear in the formalism, but the modes
at all momenta so that the important intrinsic dephasing mechanism is 
included in the theoretical treatment without adhoc assumptions and phenomenological
parameters such as the Gilbert damping. We stress that we do not include
spin-lattice relaxation in our equations in order to elucidate the effects of dephasing
alone. A Gilbert damping can be added in a further step to include relaxation; but this
is beyond the scope of the present work.

In the preceding study \cite{bolsm23} the switching was investigated
as it is induced by a constant, homogeneous magnetic field $h=g\mu_\text{B}B$ applied to
a two dimensional square lattice. This switching
field needs to be sufficiently strong to overcome the anisotropy.
At zero temperature, we found that the threshold field is fairly
precisely given by the spin gap $\Delta=h_\thr$. But at temperatures
close to the N\'eel temperature, the required threshold field becomes small.

Experimentally the samples are dealt with at temperatures
well below the ordering N\'eel temperature $T_\text{N}$. Heating up close to 
$T_\text{N}$ necessitates heating pulses which
require time and will slow down any switching. Thus, it is attractive to look
for other means to lower the minimum switching fields. Generally, overcoming
some energy barrier in quantum mechanics can be achieved by exciting
in resonance, see e.g.\  Ref.\ \cite{miyas23}.
Hence, it is promising to extend the previous treatment to time-dependent
control fields which can be tuned in resonance to the energy gap
of the magnetic system to be switched. With realizations in mind,
such control fields can be generated by THz pulses because their
frequencies lie well in the range of magnetic gaps.

In this work, we show that the switching of 
the antiferromagnetic sublattice magnetization  
can be achieved by applying experimentally accessible THz pulses with moderate amplitudes 
even at zero temperature. The optimum angular carrier
frequency of the pulses lies slightly below the spin gap $\Delta/\hbar$. The theoretical 
approach and equations stay the same as in Ref.\ \cite{bolsm23} except that the control 
magnetic field acquires a time dependence.
Furthermore, we extend previous results from two to three dimensions, i.e., we complement 
results for the square lattice by results for the cubic lattice. This important extension
becomes possible by methodological progress: generalizing the equations
from momentum space to a sort of frequency space weighted by the density-of-states.

This paper is organized as follows. In Section \ref{s:model} we briefly recall the main equations.
The magnetization dynamics of antiferromagnets under short pulses will be analyzed 
in Sect.\ \ref{s:dynamics}. Additionally, the set of optimal parameters will be established. 
Sect.\ \ref{s:conclusion} summarizes our results.

	  %==============================================================================
    % Equilibrium 
    %==============================================================================
    \section{Model and theoretical tools}
    \label{s:model}

The approach focuses on the two dimensional square and the three dimensional cubic lattices with spin $S=1/2$. The first step in the theoretical treatment is to determine the initial conditions, i.e., the thermodynamic equilibrium. This has been done in literature many times  \cite{auerb88,auerb94,silva02,bolsm23}. We recall the equilibrium below to the extent that it fixes the initial values as in Ref.\ \cite{bolsm23}. 
	
\subsection{Model and Schwinger boson mean-field theory}
	\label{ss:model1}
	
We consider a Heisenberg model with easy-axis anisotropy defined by the anisotropy parameter 
$\chi=J_{xy}/J_z\in [0,1]$ so that the Hamiltonian takes the form
\begin{equation}
      \mathcal{H}_0= J \sumnn \left\{\frac{\chi}{2}\lr{S^+_iS^-_j + S^-_iS^+_j} + S^z_i S^z_j \right\}
			\label{eqn:anizHam}
\end{equation}
in terms of the spin ladder operators.
The switching is induced by the time dependent magnetic field included in the formalism by adding the term 
 \begin{equation}\label{eqn:uniform}
    \mathcal{H}(t)= - \mathbf{h}(t)\cdot\sum_i \mathbf{S}_i .
\end{equation}
In the simulations, we choose the magnetic field to point along the $y$ axis
because we rotate sublattice B about the spin $y$ axis by 180$^\circ$ to restore translational invariance
of the original lattice even in the long-range ordered state, see below. 
In principle, the $x$ and the $y$ axis are equivalent in this problem,
but due to the sublattice rotation the application of the control field along the $y$ axis
is convenient because this preserves the translational invariance of the resulting Hamiltonian.

\begin{figure}[htb]
    \centering
    \includegraphics[width=\columnwidth]{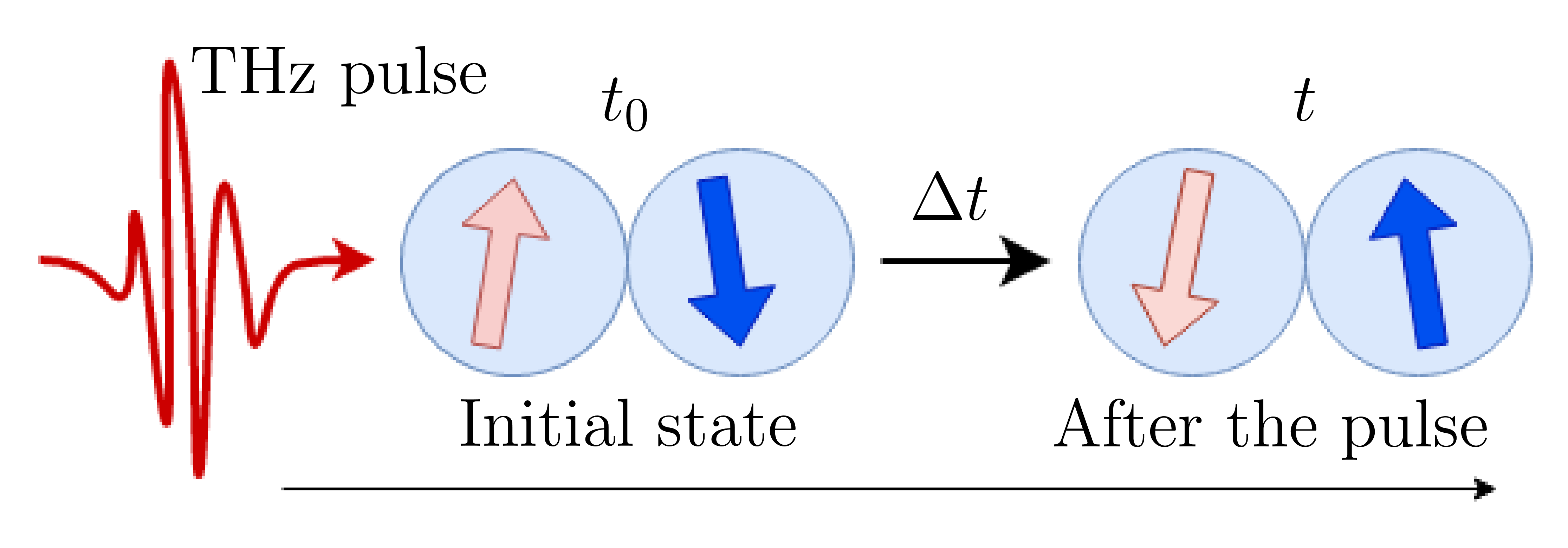}
    \caption{Illustration of the switching after a THz magnetic field pulse. 
		The delay $\Delta t=t-t_0$ is of the order of picoseconds. The magnetization of the $A$ sublattice
		is plotted as redish arrow while the magnetization of the $B$ sublattice is plotted as blue arrow.}
    \label{fig:illustrate}
\end{figure}

Figure \ref{fig:illustrate} sketches the process of switching in our case by the help of ultrashort magnetic field pulses. One of our goals is to obtain switching with $\Delta t$ being of the order of picoseconds. If this
succeeds one can claim benefits of antiferromagnets over ferromagnets.

We begin with a representation of quantum spin operators in terms of Schwinger bosons
\cite{schwi52,marde10,pires21}. Two flavors of bosons $a_i$ and $b_i$ can be present at one site 
of the lattice  and express the spin operators by
\bs
\begin{align} 
S_i^+ &= a_i^\dagger b_i
\\ 
S_i^- &= b_i^\dagger a_i
\\
S_i^z &= \frac{1}{2}\lr{\nbi{a} - \nbi{b}}.
\end{align}
\es
The bosonic Hilbert space exceeds the physical relevant subspace so that
a constraint needs to be included which reads
\be
\label{constraint}
2S= \nbi{a} + \nbi{b}.
\ee
The sublattice magnetization reads
\begin{equation}\label{eq:magnetization}
    m=\langle S_i^z \rangle = \frac{1}{2}\lr{\langle \nbi{a} \rangle  - \langle \nbi{b}\rangle},
\end{equation}	
so that the mean occupation of the bosons determines the magnetization. 
For further analysis, the bipartite lattice is divided into two interpenetrating sublattices, A and B as is 
common for unfrustrated antiferromagnets so that the nearest neighbors of the A spins are all on sublattice B 
and vice versa. Additionally, we introduce bond operators $A_{ij}=a_ib_j-b_ia_j$  and $B_{ij}=a_ib_j+b_ia_j$.
Then, it is convenient to apply a canonical transformation
to the Hamiltonian in which the spins on one sublattice, say B, are rotated by $180^\circ$ 
about the $y$ axis with the effect that $a_j \rightarrow -b_j$ and $b_j \rightarrow a_j$.   Consequently, the bond operators transform into the simpler, more symmetric form  $A_{ij}=a_ia_j+b_ib_j$  and $B_{ij}=a_ia_j-b_ib_j$.  Remarkably, the Hamiltonian keeps translational invariance in this representation even in the
antiferromagnetically ordered phase. Finally, we simplify the resulting quartic Hamiltonian
by applying a mean-field theory using the expectation values \cite{auerb94,bolsm23}
\bs
\begin{align}
 A &\coloneqq \langle A_{ij}\rangle
\\
B &\coloneqq \langle B_{ij}\rangle.
\end{align}
\es

The  total mean-field Hamiltonian $ \mathcal{H} = \mathcal{H} _0+ \mathcal{H} (t)$ from \eqref{eqn:anizHam} 
and \eqref{eqn:uniform} takes the following form after a standard Fourier transformation
\begin{align}
\nonumber
   \mathcal{H} &= E_0 -\frac{z}{8}\sumk \gam \Big(C_-\ak^\dagger \akm^\dagger + C_+\bk^\dagger\bkm^\dagger
	\\ \nonumber
	& \qquad + C_-^*\ak\akm + C_+^*\bk\bkm\Big) + \lambda \sumk \lr{\nbk{a} + \nbk{b}} 
	\\
	& \qquad - \frac{h_y(t)}{2i}\sumk\lr{\ak^\dagger\bk - \bk^\dagger\ak},
	\label{eq:hamilton-switch}
\end{align} 
where the superscript $*$ stands for the complex conjugate. The momentum dependence
enters through the parameter
\be
\label{eq:gam_def}
\gam \coloneqq \frac{1}{d} \sum_{i=1}^d \cos(k_i),
\ee
where we set the lattice constant to unity and $k_i$ is the component of the wave vector
in $i$ direction.
The coordination number, i.e., the number of nearest neighbors of each site, 
is denoted by $z$ given by twice the dimension $2d$ for hypercubic lattices and 
we use
\be
\label{eq:C-def}
C_{\pm} \coloneqq A(1+\chi)\mp B(1-\chi).
\ee
The  energy constant reads 
\be
E_0 \coloneqq \frac{zN}{4}\lrg{|A|^2 + |B|^2 +\chi(AB^*+BA^*)+2S^2}.
\ee 
The term with the site independent Lagrange multiplier  $\lambda$ has been added by hand in order
to ensure that the constraint \eqref{constraint} is fulfilled, at least on average over the whole
lattice.

In the case of equilibrium without time-dependent Zeeman term, the gapped dispersion relations can be obtained  from standard Bogoliubov transformations to diagonal $\alpha$- and $\beta$-bosons with dispersion $\omega^-_\mathbf{k}$ and $\omega^+_\mathbf{k}$, respectively, which are given by
\be
\label{dispersion}
    \omega^{\pm}_\mathbf{k} = \sqrt{\lambda^2-\lr{{z|C_\pm|\gam}/{4}}^2}.
\ee
Here, $C_\pm$ is still given  by Eq.\ \eqref{eq:C-def}, but the expectation values $A$ and $B$ and thus $C_\pm$ can be chosen real.  In the above dispersion relations, both types of bosons acquire a energy gap as 
\be
\Delta^\pm \coloneqq  \omega^\pm_{\textbf{k}=0}.
\ee
If we iterate the above equations, see Appendix \ref{a:appendixA}, to reach
convergence in the expectation values starting with more $a$ bosons
than $b$ bosons, we obtain a positive sublattice magnetization and the $b$ bosons have the
 larger energy gap $\Delta^+$. This is the initial scenario from which we start. Then, the 
physical spin gap is given by the difference 
\be
\label{eq:gap-def}
\Delta \coloneqq  \Delta^+ - \Delta^-
\ee
because the boson number in the physical subspace can only change by two, not by one.
Of course, we could also start from majority $b$-bosons; but this would only make
a difference in the sign of the sublattice magnetization.

For a static control field we found that the minimum field $h_\thr$
is almost equal to the spin gap $\Delta$.
In particular for $\chi \rightarrow 1$,  $h_\thr$ 
and $\Delta$ both vanish proportional to $\sqrt{1-\chi^2}$ \cite{bolsm23}.  
In Schwinger boson representation, the switching can be understood
as converting the majority $a$ bosons to $b$ bosons, i.e., to make
the $b$ bosons the majority bosons.
In parallel, the size of the spin gap determines the robustness 
of the system against any kind of perturbation. 
In this work, we consider oscillating fields of short pulses with angular frequencies 
close to $\Delta/\hbar$. Indeed,  such coupling in  resonance facilitates the spin reversal. 

For the time dependent mean-field theory, we need the time dependence of expectation values
which can be computed by Heisenberg's equation of motion in momentum space. They can
be efficiently expressed as functions of the parameter $\gamma$ because
the momentum dependence only enters via this parameter, see Sect.\ \ref{ss:model2}. 
The resulting differential equations \eqref{eqn:DissEQ}
are given in in Appendix \ref{a:appendixA}. Eventually, the dynamics of sublattice 
magnetization  follows from
\be
m(t) = \frac{1}{2N}\sumk\big(\lara{\nbk{a}} - \lara{\nbk{b}}\big). 
\label{eq:magdynamics}
\ee

\subsection{Equations of motion based on  the density-of-states of $\gam$}
	\label{ss:model2}

In order to calculate the quantities in the systems under study and to solve the 
differential equations \eqref{eqn:DissEQ} fulfilling Eqs.\ \eqref{eq:ABS-compute}, 
one needs to carry out sums of the 
type $\sum_\mathbf{k} F_\mathbf{k}$ where the wave vectors $\mathbf{k}$ run over 
the Brillouin zone. In the thermodynamic limit, these sums become integrals.
These integrals can be simplified further because the dependence on the wave vectors
only enters through $\gam$ defined in \eqref{eq:gam_def}, see for instance 
Eq.\ \eqref{eq:hamilton-switch}, Eq.\ \eqref{eq:ABS-compute},
or Eqs.\ \eqref{eqn:DissEQ}. Hence, one can replace the integrals over
all wave vectors by a single integral over $\gamma \in [-1,1]$. But it is 
important to weight the contributions according to how many wave vectors
contribute to a certain value of $\gamma$ \cite{auerb94,marde10}. 
This is achieved as usual
by a density-of-states $\rho_d$ in dimension $d$ 
as function of $\gamma$. We define it as
\be
\label{eq:rho-def}
\rho_d(\gamma) \coloneqq \frac{1}{(2\pi)^d} 
\int_\text{BZ} \delta(\gamma - \gam) dk^d.
\ee

With the help of $\rho(\gamma)$ 
sums over the two or three dimensional Brillouin zone 
can be reduced to one dimensional integrals in the thermodynamic limit
\begin{equation}
\label{eq:sumtoint}
    \lim_{N\to \infty} \frac{1}{N}\sum_\mathbf{k} F(\gam) =
		\int_{-1}^1d\gamma \rho(\gamma) F(\gamma).
\end{equation}
In the numerical calculations, we  discretize these integration again.
But the calculation remains very efficient because only a one dimensional
integration needs to be dealt with. In practice, we use the 
summed Newton-Cotes midpoint interval rule. We checked the results obtained via \eqref{eq:sumtoint}
on a sample basis by sums over the first Brillouin zone of the square lattice
for linear system size $L=500$  and obtained the same results within the numerical
accuracy.

    %==============================================================================
    % Dynamics
    %==============================================================================
    \section{Effect of short pulses}
    \label{s:dynamics}
		
We aim at switching the antiferromagnetic magnetization by pulses of low magnetic fields in the THz range. 
Thus, we want to model and to optimize realistic pulses so that our results can guide future experiments. 
Short magnetic pulses are technologically well possible in contrast to continuous excitation with continuous
radiation. Thus, we focus on pulses of the form
\begin{equation}
\label{shortpulse}
h_y(t)=h_0\cos(\alpha\Delta (t-3\tau)+\phi_0)\cdot e^{-\frac{(t-3\tau)^2}{2\tau^2}}  
\end{equation}
where $h_0$ is the maximum amplitude, $\Delta$ is the spin gap in Eq. \eqref{eq:gap-def} which essentially
sets the excitation frequency (setting $\hbar=1$) although it is slightly modified by the
prefactor $0 < \alpha \le 1$; a phase shift $\phi_0$ is also accounted for. 
It helps  to tune the field induced torque  in phase with the spin precessions  as discussed in Ref.\
 \cite{kampf11}. The envelope is chosen Gaussian to treat a generic pulse
of essentially finite duration $\tau$. We start our simulations at $t=0$; in order to capture
the significant part of the pulse, it is shifted by $3\tau$. 
We strive for the optimum values of the parameters in \eqref{shortpulse} for switching.
In Appendix \ref{a:appendixB} we provide results also for two other pulse shapes.

	\subsection{Dynamics of the magnetization for the square lattice at zero temperature}
	\label{ss:dynamicsatzero}	

\begin{figure}[htb]
    \centering
 \includegraphics[width=\columnwidth]{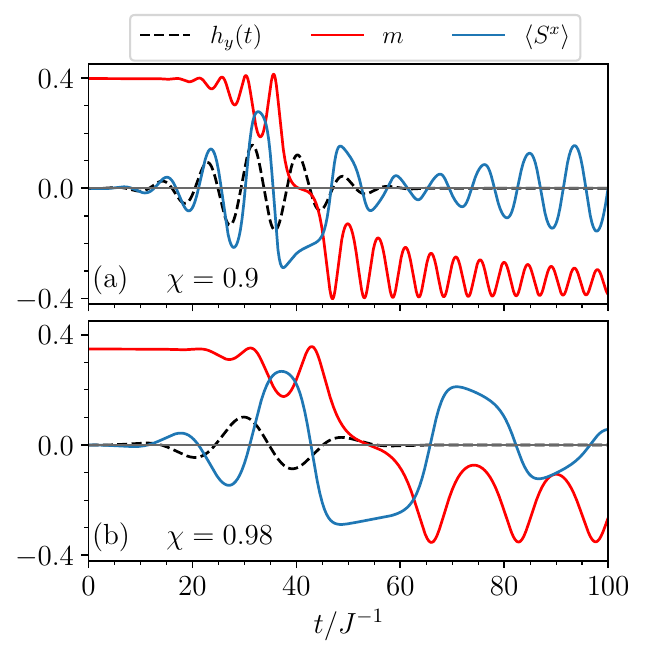}
    \caption{Temporal evolution of the sublattice magnetization $m(t)$ (red curve) for $\chi=0.9$,  $h_0=0.158 \, J$ in panel (a) and $\chi=0.98$,  $h_0=0.107 \, J$ in panel (b).  The evolution of the pulse $h_y(t)$ is included as dashed black curves. 
		The dynamics of the transversal spin component $\langle S^x \rangle$ is  shown in light blue.
The spin gap takes the values $\Delta (\chi=0.9)=0.837 \, J$ and $\Delta (\chi=0.98)=0.357 \, J$.
Other parameters are chosen as $\alpha=0.85$, $\phi_0=\pi/3$, and $\tau=10 \, J^{-1}$.}
    \label{fig:gaps-mag}
\end{figure}

Figure \ref{fig:gaps-mag} demonstrates how the magnetization evolves 
in time as a result of the pulse in Eq. \eqref{shortpulse}. 
We emphasize that a  single sign change of the magnetization is
indeed realized  although the magnetization continues to oscillate. The amplitude
of these oscillations decreases due to dephasing 
so that on the long run the stationary opposite magnetization is realized. 
Thus, we take such single sign change a signature of a switch of magnetization. 
A purely classical treatment based on Landau-Lifshitz equations alone would
yield infinitely lasting oscillations without Gilbert damping, see Appendix \ref{a:appendixD}. 
This underlines the relevance of capturing dephasing by treating all modes of a spin wave
theory. Of course, the dephasing oscillations would decay even faster if 
spin-lattice relaxation were included additionally, for instance by Gilbert damping.

The magnetic amplitudes $h_0=0.158 \, J$ and $h_0=0.107 \, J$ are slightly higher than
the dynamic threshold fields for $\chi=0.9$ given by $h_0^\thr=0.1578 \, J$ 
and for $\chi=0.98$ given by $h_0^\thr=0.1065 \, J$.
For comparison, the threshold fields required in the static case are
$h_\thr=0.863 \, J$ and  $h_\thr=0.364 \, J$, correspondingly. 
One sees that the control at resonance frequencies truly pays.
The black dashed curves depicting $h_y(t)$ clearly show quicker oscillations for a larger
spin gap comparing Fig.\ \ref{fig:gaps-mag}(a) and in Fig.\ \ref{fig:gaps-mag}(b). 
Another notable point is that the switching takes a bit longer for lower anisotropy
with lower spin gap. Still, switching takes place on a THz scale in both cases. 
If we consider a realistic antiferromagnetic exchange coupling of $J\approx 10 \,$meV, 
the switching occurs at $t=40.5 \,J^{-1}\approx 2.63 \,$ps for $\chi=0.9$ and  
$t=53.7 \,J^{-1}\approx 3.53 \,$ps for $\chi=0.98$.
In parallel, the oscillations of the transversal component 
$\langle S^x \rangle$ is also slower close to isotropy because the spin gap is smaller
and thus also the chosen pulse frequency. 
Based on these results, we emphasize that the switching is obtained with relatively low maximum
amplitudes of the THz pulse. The other parameters are chosen as $\alpha=0.85$, $\tau=10$ $J^{-1}$,
 and $\phi_0=\pi/3$ because our analyses indicate that these values are best suited for switching,
for further discussion see below.

\begin{figure}[htb]
    \centering
    \includegraphics[width=\columnwidth]{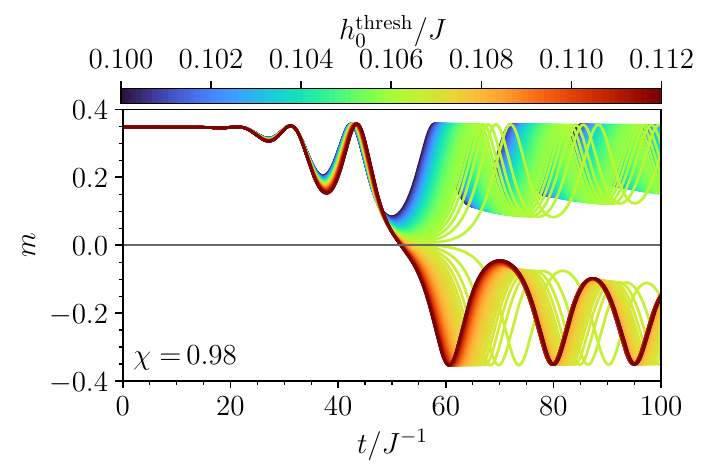}
    \caption{Dynamics of the magnetization $m(t)$ at $\chi=0.98$ for $h_0\in[0.100\,J,\,0.112\,J]$. 
		For amplitudes above $h_0^\thr=0.1065 \, J$ switching takes place. 
		All other parameters are the same as in 
		Fig.\ \ref{fig:gaps-mag}.}
    \label{fig:dynamics}
\end{figure}

To show the dynamics of the magnetization for various amplitudes, we plot it in Fig.\ \ref{fig:dynamics}. 
This plot illustrates nicely the threshold behavior which occurs also for the time-dependent pulses.
Only for maximal amplitudes above a certain value the magnetization switches sign;
this value is about $h_0^\thr \approx 0.1065 $ $J$. For lower amplitudes no sign reversal takes place. 

\begin{figure}[htb]
    \centering
    \includegraphics[width=\columnwidth]{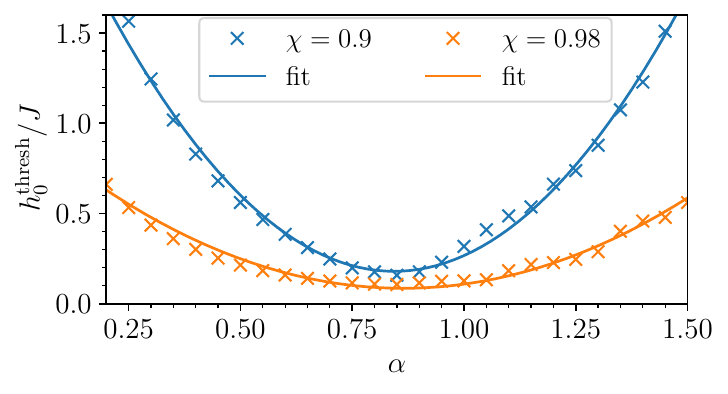}
    \caption{Threshold value $h_0^\thr$ as function of the renormalization $\alpha$ of the excitation frequency in \eqref{shortpulse} for two anisotropies. 
		The minimas are found by fitting parabolae. Clearly, the optimum excitation
		is obtained for $\alpha \approx 0.85$.
		The other parameters of the pulse are $\tau=10 \,J^{-1}$,  $\phi_0=\pi/3$.}
    \label{fig:alpha}
\end{figure}

In Fig.\ \ref{fig:alpha} one can see that the excitation frequency  plays an important role in switching. 
Generically, it helps if the pulse is in resonance with the spin gap $\Delta$. But the activation
energy induced by the anisotropy  acts like a hill which needs to be overcome. Thus, in the course of
switching the slope of this hill diminishes and the required frequency diminishes as well. This
has been shown explicitly for a large spin \cite{miyas23}. To account for this effect, at least partially,
we introduced the frequency renormalization $\alpha$ in the pulse \eqref{shortpulse}. Indeed,
the data depicted in Fig.\ \ref{fig:alpha} shows that the lowest threshold field is not obtained for
$\alpha=1$, but for values around $0.8$. In order to determine the minimum we fitted the data
by parabolae and obtained for both anisotropies an optimum value of $\alpha\approx 0.85$.

Another parameter in the ansatz \eqref{shortpulse} is $\tau$. It appears as delay and 
as pulse duration. We refrain from considering the delay separately because for a Gaussian pulse 
one can be quite sure that it has essentially all its effect if it is applied over
a full-width interval of six times its standard deviation which is given by $\tau$ in \eqref{shortpulse}.
Thus, delaying the pulse peak to $3\tau$ captures almost all the effect of the pulse.

\begin{figure}[htb]
    \centering
    \includegraphics[width=\columnwidth]{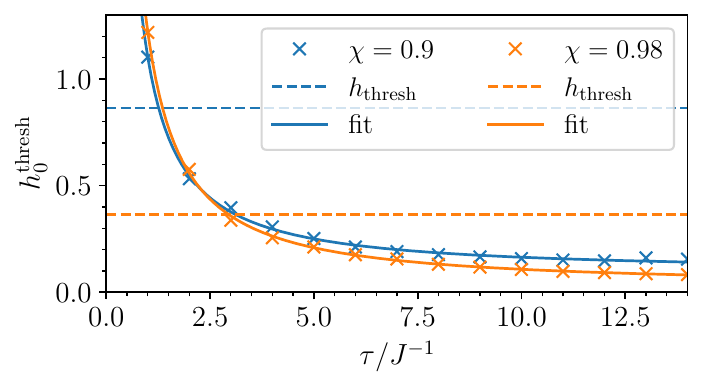}
    \caption{Threshold fields (symbols) vs.\ pulse duration $\tau$ including the indicated fits. The other parameters are $\alpha=0.85$ and $\phi_0=\pi/3$. 
Dashed horizontal lines correspond to the threshold values for uniform magnetic field with $h_\thr=0.863 \,J$ and 
$h_\thr=0.364 \,J$ for $\chi=0.9$ and $\chi=0.98$, respectively. The fits (solid lines) are power laws
 $h_0^\thr=a\tau^b+c$ with parameters $a=1.007 \, J^{b+1}$, $b=-1.158$, $c=0.093 \, J$ for $\chi=0.9$ and $a=1.195  \, J^{b+1}$, $b=-1.167$, $c=0.026 \, J$ for $\chi=0.98$. }
    \label{fig:tau2D}
\end{figure}

Still, the duration of the pulse matters so that we analyze the dependence 
on $\tau$ in Fig.\ \ref{fig:tau2D}. Obviously, the threshold field is very high for small values of $\tau$.
For increasing $\tau$ the threshold field decreases as one would naively expect
because a longer lasting pulse has more effect or it can have the same effect at lower
amplitude. We fit the threshold fields by a power law $a \tau^b +c$ and find the values
indicated in the caption. The exponent is of the order of $-1$, but the finite values $c>0$ 
indicates that even a very long pulse requires some finite minimum amplitude
to achieve switching. This is in line with what we found for static fields \cite{bolsm23}.
The dashed horizontal lines indicate the threshold fields required for
static control fields. From the data in Fig.\ \ref{fig:tau2D} we choose $\tau=10/J$
as a reasonable tradeoff between a small threshold amplitude and a not too long
switching time of about $40/J$.

\begin{figure}[htb]
    \centering
    \includegraphics[width=\columnwidth]{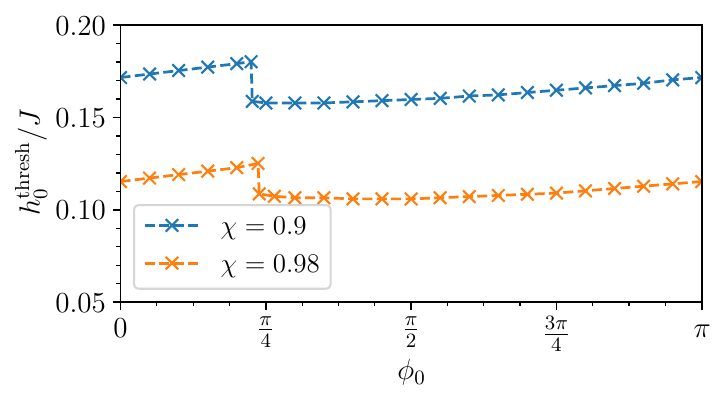}
    \caption{Threshold fields (symbols) vs.\ phase $\phi_0$ in (\ref{shortpulse}) for 
		$\alpha=0.85$ and $\tau=10$ $ J^{-1}$.}
    \label{fig:phase}
\end{figure}

Last, but not least, we study the effect of the phase on switching. Figure \ref{fig:phase} shows the threshold 
amplitude of the pulse as function of the phase $\phi_0$. The overall dependence is rather weak, i.e.,
the phase does not have a large effect on the threshold field. Quite unexpectedly, however, there
is one jump occuring where the threshold field suddenly drops upon changing the phase. 
This jump is related to a sudden shift in time at which the switching actually takes place.
Just below a certain phase $\phi_0$ the switching occurs at a zero of the control field while
just above that value the switching occurs almost a control cycle later. For an illustration of
this subtlety we refer the reader to Fig.\ \ref{phaseshift} in Appendix \ref{a:appendixB}.
We have chosen the near optimum value of $\phi_0 = \pi/3$; it is applicable 
for any value of a weak anisotropy parameter in our analysis since the influence of the
phase is rather weak.

\subsection{Dynamics of the magnetization for the cubic lattice at zero temperature}
	\label{ss:dynamics3D}

\begin{figure}[htb]
    \centering
    \includegraphics[width=\columnwidth]{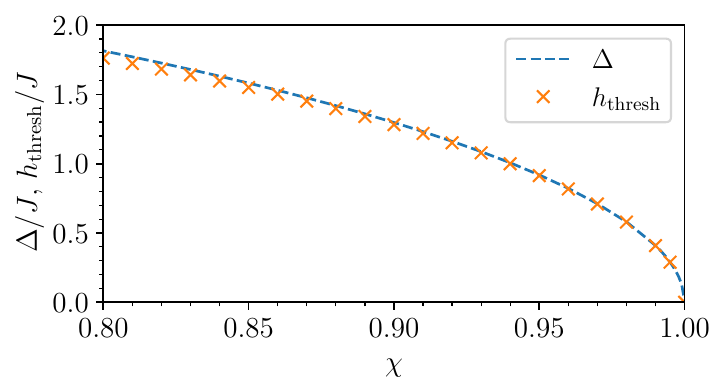}
    \caption{Threshold field $h_\thr$ and spin gap $\Delta$ both at zero 
		temperature as function of the anisotropy $\chi$. They almost coincide 
		as was found previously in two dimensions
		as well \cite{bolsm23}.}
    \label{fig:thresh3D}
\end{figure}

Since we solve the equations of motion not in reciprocal space, but in frequency space, more 
precisely in $\gamma$ space, there is no major difference in the numerical effort
between treating the two dimensional and the three dimensional cases. The only necessary
modification is to choose the appropriate density-of-states $\rho_d$. Explicit results
can be taken from literature \cite{hanis97}. As a first step, we compute and compare
the spin gap and the static threshold field at zero temperature in Fig.\ \ref{fig:thresh3D}.
As for the square lattice, both quantities agree very well. This supports the view
that the spin gap is a measure for the activation energy to be overcome for switching
in case of anisotropies.

\begin{figure}[htb]
    \centering
    \includegraphics[width=\columnwidth]{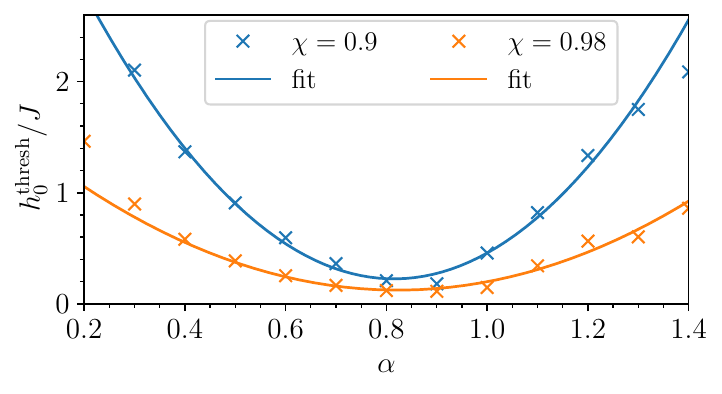}
    \caption{The threshold value $h_0^\thr$ 
		as function of the renormalization $\alpha$ of the excitation frequency in \eqref{shortpulse} 
		for two anisotropies. The minima are found by fitting parabolae. The other parameters of the pulse 
		are $\tau=10 \,J^{-1}$,  $\phi_0=\pi/3$.}
    \label{fig:alpha3D}
\end{figure}

Next, we proceed with the analysis of the effects of short Gaussian pulses \eqref{shortpulse}.
First, we determine the optimum value of the frequency renormalization $\alpha$ in Fig.\ 
\ref{fig:alpha3D}. The results are very similar to the results in two dimensions.
A slight lowering of the spin gap frequency by $\alpha \approx 0.85$ appears to
allow for the lowest amplitudes for switching.

Also the dependence of the threshold amplitude $h^\thr_0$ on the pulse duration $\tau$ is 
qualitatively the same as for the square lattice. The only quantitative difference is the value
 of the threshold amplitude $h_0^\thr$ which is higher for the cubic lattice in line with
the spin gap. This is a direct consequence of the higher coordination number stabilizing
the ordered phase and enhancing the effect of the exchange anisotropy. We conclude that the theoretical findings  will carry over to many other comparable types of lattices.

\begin{figure}[htb]
    \centering
    \includegraphics[width=\columnwidth]{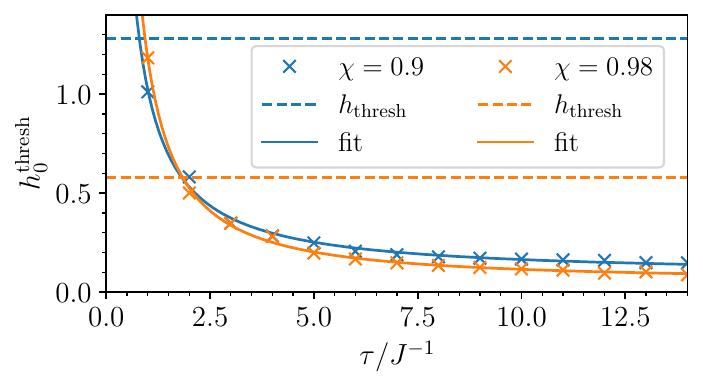}
    \caption{Threshold fields (symbols) vs.\ pulse duration $\tau$ including the indicated fits. The other parameters are  $\alpha=0.85$ and $\phi_0=\pi/3$. 
Dashed horizontal lines correspond to threshold values at uniform magnetic field with $h_\thr=1.281 \,J$ 
and $h_\thr=0.578 \,J$ for $\chi=0.9$ and $\chi=0.98$, respectively. Fitting is done with the same power law $h_0^\thr=a\tau^b+c$ as in 2D. The fitting parameters are $a=0.937 \, J^{b+1}$, $b=-1.068$, $c=0.084 \, J$ for $\chi=0.9$ and $a=1.127 \, J^{b+1}$, $b=-1.248$, $c=0.051 \, J$ for $\chi=0.98$. }
    \label{fig:tau3D}
\end{figure}

\begin{figure}[htb]
    \centering
    \includegraphics[width=\columnwidth]{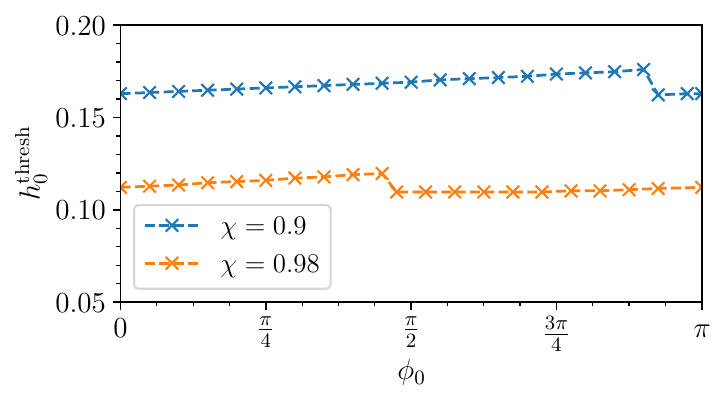}
    \caption{Threshold fields (symbols) vs.\ phase $\phi_0$ in (\ref{shortpulse}) for 
		$\alpha=0.85$ and $\tau=10$ $ J^{-1}$.}
    \label{fig:phi3D}
\end{figure}

Figure \ref{fig:phi3D} shows the threshold field dependence on the phase in three dimensions.
Again, we find that the phase of the control field has only a weak effect on the threshold value as for the square lattice. We again obtained a jump in the phase dependence resulting from a shift in the switching time
 as shown and discussed in Appendix \ref{a:appendixB} for the square lattice.
Due to the small influence of the precise phase value, we fixed the  phase to the value $\phi_0=\pi/3$ for cubic lattice although this does not represent the best possible value, cf.\ Fig.\ \ref{fig:phi3D},
but the influence is small.

	\subsection{Dynamics at finite temperature}
	\label{ss:dynamicsatfinite}

\begin{figure}[htb]
    \centering
    \includegraphics[width=\columnwidth]{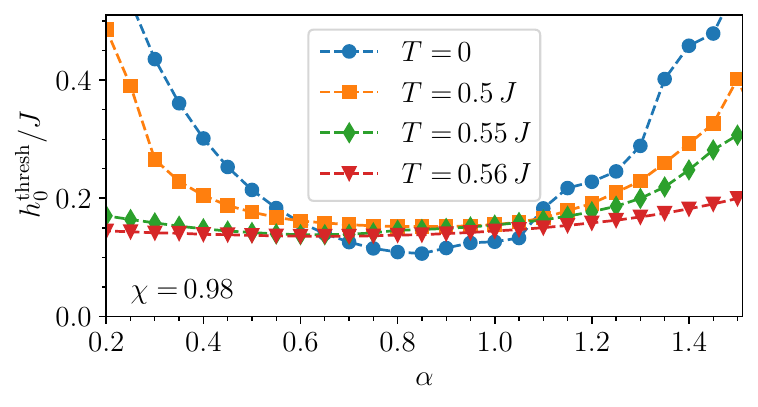}
    \caption{The threshold value $h_0^\thr$ as function of the renormalization $\alpha$ of the excitation frequency in \eqref{shortpulse} as in Fig.\ \ref{fig:alpha} for $\chi=0.98$ for various temperatures.
		Recall that the N\'eel temperature takes the value $T_\text{N}=0.578 \,J$ \cite{bolsm23} which is between $T_\text{N}=70\,$K and   $700 \,$K for realistic exchange couplings.
		All other parameters are chosen as before.}
    \label{fig:fintemp}
\end{figure}

	In the case of switching with static fields it was found that elevated temperatures
	in the vicinity of the N\'eel temperature reduced the threshold field $h_\thr$ significantly \cite{bolsm23}.
	Thus, it suggests itself to study the effect of finite temperatures also for time-dependent short
	pulses. This is achieved in Fig.\ \ref{fig:fintemp} for the square lattice where
	the threshold amplitudes  are depicted as function of  $\alpha$ for $\chi=0.98$ for four different temperatures. 
	Interestingly enough, the overall shape of this dependence becomes flatter and flatter as
	the temperature is increased. This implies that for control pulses far off resonance the 
	increased temperature facilitates switching as found before in the static case \cite{bolsm23}.
	But in the optimum region around $\alpha=0.85$ the changes are very small. 
	This is an interesting finding carrying the message to experiment that 
	it pays off better to choose the optimum resonance conditions, perhaps accounting
	for some renormalization by 10 to 20\%, than to work at elevated temperatures.
	If the N\'eel temperature lies well above room temperature, all experimental
	studies can well be conducted at room temperature. For instance, 
	La$_2$CuO$_4$ ($T_{\mathrm{Neel}}=325\,$K) and YBa$_2$Cu$_3$O$_{6.15}$ ($T_{\mathrm{Neel}}=410\,$K) are  representatives of $S=1/2$ systems which are mostly two-dimensional, but with a certain
	perpendicular coupling warranting a phase transition at finite temperatures.

    \section{Conclusions}
    \label{s:conclusion}
		
We studied the time-dependent control of the sublattice magnetization in 
two and three dimensional anisotropic quantum antiferromagnets. 
The motivation is to use such systems for data storage; the anisotropy
being crucial to make the long-range order robust.
Our study uses spin wave theory
based on the representation of the spins by Schwinger bosons and the corresponding
mean-field theory. The representation by Schwinger bosons has the fundamental advantage 		
that it can capture any orientation of the magnetization and vary it in time without
losing its justification. In this way, magnon modes at \emph{all} wave vectors ${\bf k}$ are
included in the description, not only the macrospins corresponding to ${\bf k}={\bf 0}$
and ${\bf k}=(\pi,\pi)^\top$ or ${\bf k}=(\pi,\pi,\pi)^\top$, respectively.
Thereby, the intrinsic dephasing effects are included and no Gilbert damping
needs to be introduced to account for it. We point out, that spin-lattice relaxation is
not included. In order to account for such a relaxation channel a Gilbert damping term
can be taken into account in future work.

The main focus was to study short pulses as they can be generated nowadays
in the THz regime. As expected such pulses are most effective if their
carrier angular frequency is chosen to be in the range of the spin gap $\Delta/\hbar$.
But we found that the optimum effect is obtained for a frequency lowered by about 15\%.
We attribute this finding to the reduction of the spin gap in the course
of switching similar to the activation energy of a large anisotropic spin \cite{miyas23}.
We found that a short Gaussian pulse is a very suitable choice to achieve
switching of the magnetization fairly quickly. Unexpectedly, finite temperature
does not help much to improve the switching efficiency if the pulses have been 
optimized.

Other choices of pulse shapes, for instance
a pulse shaped like a sinus cardinalis \cite{miyas23}, require stronger fields.
A continuous oscillating control field can switch the magnetization with 
low amplitudes, but it takes fairly long for the switching to take place.
Also the Gaussian pulses require less maximum amplitude if they are applied
longer, i.e., if the pulse duration is larger. On the downside, 
this makes the switching process slower. 

A quantitative estimate strongly depends on the numbers. If we assume $J\approx 10$\,meV
and $\chi=0.98$ the required Gaussian amplitude for $\tau=10/J\approx 1$\,ps is as large
as about 10\,T. If the pulse is applied about four times longer and the anisotropy is smaller, let us say
$\chi=0.995$, then the estimate yields about 1\,T. Still, these are fairly large values so that
further reductions are desirable.

A promising route consists in the use of alternating (N\'eel) effective magnetic fields. Theoretical
treatments based on macrospin theory \cite{gomon10,roy16} and recent experimental 
evidence \cite{behov23} show that the switching can be obtained efficiently 
by exploiting the exchange field induced by the magnetization of coupled adjacent spins.
Thus, the extension of the present treatment to
alternating N\'eel fields appears attractive.
Furthermore, it is easy to conceive further important extensions such as the treatment
of other lattices in two and three dimensions and/or higher spins $S>1/2$.
Such studies will decisively help to introduce antiferromagnets as systems for data storage
and manipulation.

    %==============================================================================
    % Acknowledgments 
    %==============================================================================
\begin{acknowledgments} 
 We are grateful to Timo Gr\"a\ss{}er, Dag-Bj\"orn Hering, 
and Joshua Alth\"user for very useful discussions. This work has been financially supported by the 
Deutsche Forschungsgemeinschaft (German Research Foundation) in project UH 90/14-1
and by the Stiftung Mercator in project Ko-2021-0027.
\end{acknowledgments}

%\bibliography{liter10.bib}


   %==============================================================================
   % Appendix 
   %==============================================================================

    \begin{appendix} 
		
\section{Dynamics of expectation values}
\label{a:appendixA} 

First, we give the expectation values in equilibrium at temperature $T$.
They provide the initial values for the subsequent dynamic calculation.
The thermal equilibrium results from standard bosonic statistics
after diagonalization of the mean-field Hamiltonian \eqref{eq:hamilton-switch} 
for $h_y=0$ by
the appropriate Bogoliubov transformation \cite{bolsm23}.
\bs
\label{eqn:exp-values}
\begin{align}
\lara{\ak\akm}_{\gamma} &= \frac{z\gamma C_-}{8\omega_k^-(\gamma)}
\coth(\frac{\omega_k^-(\gamma)}{2T}) ,\label{bbkg} 
\\ 
\lara{\bk\bkm}_{\gamma} &= \frac{z\gamma C_+}{8\omega_k^+(\gamma)}
\coth(\frac{\omega_k^+(\gamma)}{2T})   ,\label{aakg}
\\
\lara{\nbk{a}}_{\gamma} &= \frac{\lambda}{2\omega_k^-(\gamma)}
\coth(\frac{\omega_k^-(\gamma)}{2T})-\frac{1}{2}   , \label{nakg} 
\\
\lara{\nbk{b}}_{\gamma} &=  \frac{\lambda}{2\omega_k^+(\gamma)}
\coth(\frac{\omega_k^+(\gamma)}{2T})-\frac{1}{2}  ,\label{nbkg}.
\end{align}
\es
The constants $C_\pm$ are defined in Es.\ \eqref{eq:C-def} and the dispersions in
Eq.\ \eqref{dispersion}. The variables $A$ and $B$ required
to compute $C_\pm$ are defined by
\bs
\label{eq:ABS-compute}
\begin{align}
 A &= \lara{a_ia_j} + \lara{b_ib_j}
\\
 &=\int_{-1}^1 \gamma\rho_d(\gamma)\lr{\lara{\ak\akm}_{\gamma}
+\lara{\bk\bkm}_{\gamma}}d\gamma , \label{eqn:IV3} 
\\ 
 B &= \lara{a_ia_j} - \lara{b_ib_j} 
\\
&=\int_{-1}^1 \gamma\rho_d (\gamma)\lr{\lara{\ak\akm}_{\gamma}-\lara{\bk\bkm}_{\gamma}}d\gamma  ,\label{eqn:IV4}
\\
2S &= \lara{\nbi{a}} + \lara{\nbi{b}}
\\
&= \int_{-1}^1\rho_d(\gamma) \big(\lara{\nbk{a}}_{\gamma} 
+ \lara{\nbk{b}}_{\gamma}\big)d\gamma  ,\label{eqn:IV5}
\end{align}
\es
where the last equation is the average constraint on the number of
Schwinger bosons which needs to be fulfilled as well by adjusting
the Lagrange multiplier $\lambda$.
The density $\rho_d(\gamma)$ is the density-of-states in $d$ dimensions 
for $\gam$ in Eq.\ \eqref{eq:rho-def}, see Ref.\ \cite{hanis97} 
for concrete expressions in two and three dimensions.

Finally, the temporal evoluation is determined from the equations
of motion for the introduced expectation values. 
As stated before, this dynamics only depends on the value $\gam=\gamma$
\bs
\label{eqn:DissEQ}
\begin{align}
    \partial_t \lara{\ak^\dagger \ak}_{\gamma} &= -i\frac{z}{4} 
		\gamma \big(C_-^*\lara{\ak\akm}_{\gamma}-
		C_-\lara{\ak^\dagger\akm^\dagger}_{\gamma}\big) 
		\nonumber \\
	 & \quad + \frac{h_y(t)}{2}\big(\lara{\ak^\dagger\bk}_{\gamma}+
		\lara{\bk^\dagger\ak}_{\gamma}\big), \label{eqn:DissEQ1}
\\
    \partial_t \lara{\bk^\dagger \bk}_{\gamma} &=-i \frac{z}{4} 
		\gamma \big(C_+^*\lara{\bk\bkm}_{\gamma}
		-C_+\lara{\bk^\dagger\bkm^\dagger}_{\gamma}\big) 
		\nonumber \\ 
    &\quad - \frac{h_y(t)}{2}\big(\lara{\ak^\dagger\bk}_{\gamma}+
		\lara{\bk^\dagger\ak}_{\gamma}\big), \label{eqn:DissEQ2}
\\
    \partial_t \lara{\ak^\dagger \bk}_{\gamma} &= - i \frac{z}{4}\gamma 
		\big(C_-^*\lara{\ak\bkm}_{\gamma} 
		- C_+\lara{\ak^\dagger\bkm^\dagger}_{\gamma}\big)
		\nonumber \\
		&\quad+ \frac{h_y(t)}{2}\big(\lara{\bk^\dagger\bk}_{\gamma}
		-\lara{\ak^\dagger\ak}_{\gamma}\big) ,\label{eqn:DissEQ3}
\\
    \partial_t \lara{\ak \akm}_{\gamma} &=  i \frac{z}{4}\gamma 
		\big[C_- (2\lara{\nbk{a}}_{\gamma}+1)\big]
		\nonumber \\
		&\quad - 2 \lambda  i \lara{\ak\akm}_{\gamma} 
		+ h_y(t)\lara{\ak\bkm}_{\gamma} ,\label{eqn:DissEQ4}
\\
    \partial_t \lara{\bk \bkm}_{\gamma} &=  i\frac{z}{4} \gamma 
		\big[C_+ (2\lara{\nbk{b}}_{\gamma}+1)\big]
		\nonumber \\
		&\quad - 2 \lambda  i \lara{\bk\bkm}_{\gamma} 
		- h_y(t)\lara{\ak\bkm}_{\gamma} ,\label{eqn:DissEQ5}
\\
    \partial_t \lara{\ak \bkm}_{\gamma} &=  i \frac{z}{4} \gamma 
		\big( C_-\lara{\ak^\dagger\bk}_{\gamma} + 
		C_+ \lara{\bk^\dagger\ak}_{\gamma} \big)
		\nonumber \\ \nonumber
     &\quad -2 \lambda  i \lara{\ak\bkm}_{\gamma}\\
		 &\quad - \frac{h_y(t)}{2}\big(\lara{\ak\akm}_{\gamma}-\lara{\bk\bkm}_{\gamma}\big).
				\label{eqn:DissEQ6}
\end{align}
\es
This concludes the set of required equations.

\section{Sudden shift in switching time due to a change in phase}
\label{a:appendixB} 

\begin{figure}[htb]
    \centering
    \includegraphics[width=\columnwidth]{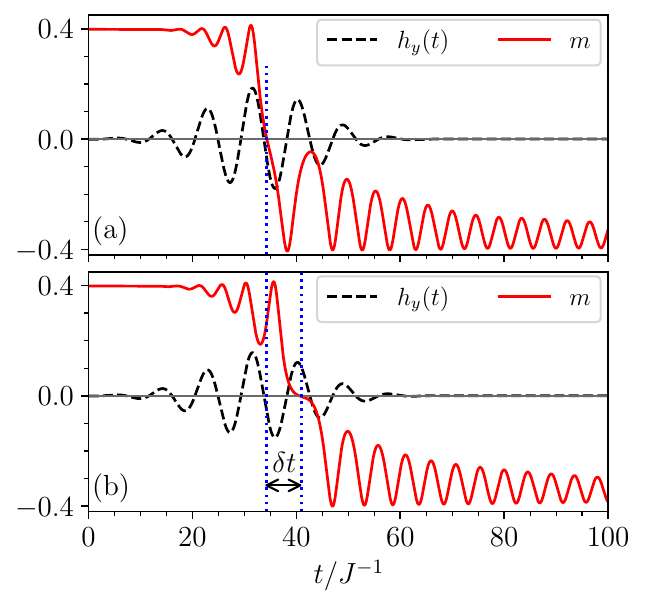}
    \caption{Evolution of the magnetization for phases just before the jump with $h_0=0.187 \,J$, $\phi_0=0.22\pi$ (a) and just after the jump with $h_0=0.158 \,J$, $\phi_0=0.23\pi$ (b) in Fig.\ \ref{fig:phase} for $\chi=0.9$.  The time shift where switching occurs due to the minute change in phase
		is denoted $\delta t$.}
    \label{phaseshift}
\end{figure}	

We emphasize that the phase of the Gaussian pulse only has a weak effect on the required field strength. 
Still, we found a  small jump in Fig.\ \ref{fig:phase}. To understand its origin, we provide 
two panels in Fig.\ \ref{phaseshift} to show the dynamics of the magnetization with phases just before and 
just after the jump. 
One can see that the slightly higher phase with a lower value of the control field allows for a switching
at a later instant in time, shifted by $\delta t$. The instant of switching occurs about 3/4 of a pulse cycle
later. Hence, the pulse takes a little  longer to induce the sign flip of $m$ which in return
allows for the lower amplitude.

	\section{Effect of alternative control fields in 2D}
	\label{a:appendixC} 

\subsection{Equally weighted superposition of frequencies}
	\label{a:longrun}	
	
\begin{figure}[htb]
    \centering
    \includegraphics[width=\columnwidth]{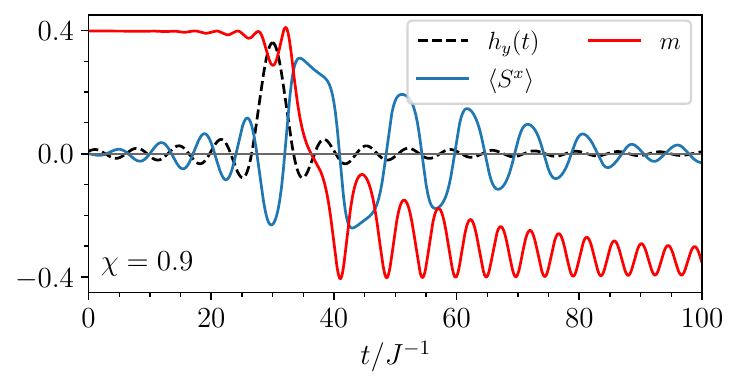}
    \caption{Evolution of the magnetization $m(t)$ and the expectation value 
		$\langle S^x\rangle$ induced by the pulse  Eq.\ \eqref{justsinpulse}
		for $h_0=0.36 \, J$, $\alpha=1.1$ and $\tau=30 \, J^{-1}$ at zero temperature for the 
		square lattice.}
    \label{sint}
\end{figure}

A typical energy landscape with a parabolic activitation maximum between the energy minima
has been studied recently by Miyashita and Barbara \cite{miyas23}. They showed that
it is advantageous to design the pulse such that in its Fourier transform 
the frequencies of all necessary level transitions are comprised with equal weight.
If one generalizes this to an infinite spin the Fourier sum becomes an integral
and the resulting pulse shape is a sinus cardinalis well-known from the
diffraction of a single slit; it reads
\begin{equation}
\label{justsinpulse}
h_y(t)=h_0\frac{\sin(\alpha\Delta (t-\tau))}{\alpha\Delta(t-\tau)}.
\end{equation}
Thus, we studied this pulse shape as well. Figure \ \ref{sint} displays a representative
result for the evolution of the magnetization for the chosen anisotropy.

Indeed, this pulse as well can achieve switching of the magnetization. On the one hand,
the required amplitude $h_0$ is lower than the threshold field for a static control
field $h_\thr=0.863\, J$ from Ref.\ \onlinecite{bolsm23}. On the other hand, 
the amplitude needed for switching is significantly larger than the one
for the optimized Gaussian pulse \eqref{shortpulse} $h_0^\thr= 0.1578\,J$,
 cf.\ Fig.\ \ref{fig:gaps-mag}.
Thus, we conclude that the pulse shape \eqref{justsinpulse} is not optimum. In addition, it is 
experimentally not easily realizable where short pulses consist 
of very few cycles only.

\begin{figure}[htb]
    \centering
    \includegraphics[width=\columnwidth]{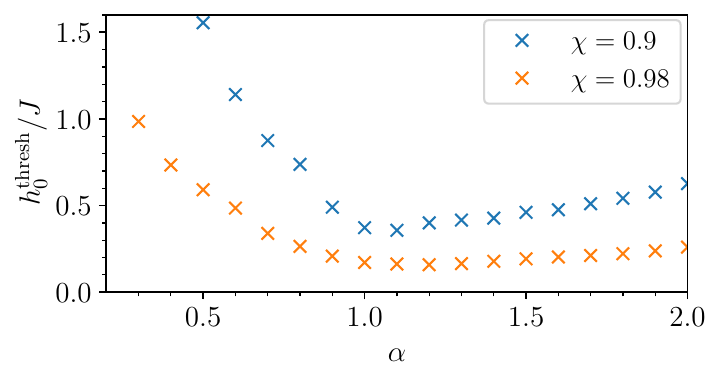}
    \caption{The threshold field dependence on $\alpha$ for 
		pulse \eqref{justsinpulse}. Here, the delay is chosen to be $\tau=30 \,J^{-1}$.}
    \label{sint_freq}
\end{figure}

For completeness, Fig.\ \refeq{sint_freq} shows  the threshold field dependence on the frequency renormalization
 $\alpha$ in \eqref{justsinpulse}. One can see that there is almost no shift away from the
resonance condition $\alpha=1$ resonance in contrast to the  Gaussian pulse in \eqref{shortpulse}.
It appears that the optimum $\alpha$ lies a bit above unity; for instance, 
we found $\alpha \approx 1.1$ with $h_0^\thr=0.357 \,J$ for $\chi=0.9$.
We attribute the finding that $\alpha <1$ is not optimum  to the fact that the pulse shape
\eqref{justsinpulse} already comprises all frequencies below $\alpha\Delta$ in its Fourier transform. 
Thus, no additional lowering is needed.

	\subsection{Continuous oscillating control field}
	\label{a:decay}	

Another conceivable pulse applies a continuous electromagnetic
wave to the sample. Within the framework of our model this corresponds to the application of
\begin{equation}
\label{justcospulse}
h_y(t)=h_0\cos(\alpha\Delta t).
\end{equation}
\begin{figure}[htb]
    \centering
    \includegraphics[width=\columnwidth]{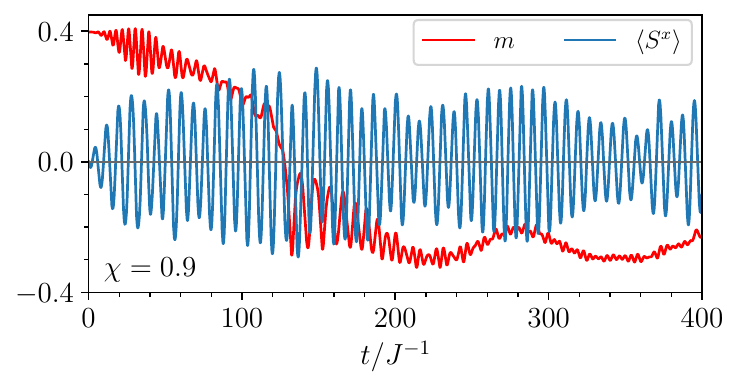}
    \caption{Switching in square lattice by long lasting control field with amplitude $h_0=0.055 \,J$ in \eqref{justcospulse}, 
		$\Delta=\Delta (\chi=0.9)$ and $\alpha=0.95$ .}
    \label{cost}
\end{figure}

\begin{figure}[htb]
    \centering
    \includegraphics[width=\columnwidth]{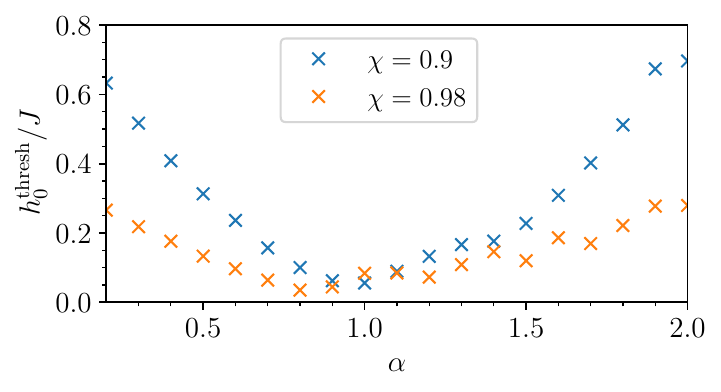}
    \caption{The threshold value $h_0^\thr$ dependence on renormalization $\alpha$ of the continuous oscillating field in \eqref{justcospulse}.}
    \label{cost:alpha}
\end{figure}

Figure  \ref{cost} displays a generic result for the continuous driving \eqref{justcospulse}
which illustrates successful switching for a fairly small amplitude, even lower than the threshold values
given in Fig.\ \ref{fig:tau2D}. But the switching takes very long so that this choice of
control does not appear appealing in practice, besides its difficulty in realization. 

For completeness, we also analyzed the effect of a renormalization $\alpha$ in Fig.\ \ref{cost:alpha}.
The effect is the similar to the effect for the  Gaussian pulse \eqref{shortpulse}. 
A value $\alpha\lessapprox 1$ allows us to reduce the amplitude slightly. But the overall
dependence on $\alpha$ is not very regular and deviates from a parabolic shape even around
its minimum. We refrain from a more detailed analysis since the continuous driving
is not easily accessible experimentally and our findings indicate unnecessary long switching
times.

\section{Comparison of the quantum and the classical behavior after a pulse}
\label{a:appendixD} 

One may wonder how strong the effect of quantum fluctuations is compared to a classical approach.
A first answer to this question lies in the initial conditions. If we deal with a small spin
such as $S=1/2$ the quantum fluctuations imply that already the initial conditions are influenced
by quantum fluctations. The initial sublattice magnetization does not take the value $\pm S$, but a 
reduced value, even at zero temperature. Yet this effect may not be too large, in particular
for large spin and dimension. In addition, one can, of course, account for the reduced initial value
in a classical simulation.

Here, we want to illustrate briefly the main difference between a spin wave calculation 
and a classical simulation. We focus on zero temperature in order to avoid to have to
distinguish between quantum and thermal fluctuations. Classically, we only need to
track the two magnetizations on the two  sublattices, i.e.,
\bs
\label{a:LL}
\begin{align}
\dot{\bm{m}}_A&=zJ(\underline{\bm{D}}\, \bm{m}_B)\times \bm{m}_A+\bm h(t)\times \bm m_A , \label{eq:classeqma}\\
\dot{\bm{m}}_B&=zJ(\underline{\bm{D}}\, \bm{m}_A)\times \bm{m}_B+\bm h(t)\times \bm m_B , \label{eq:classeqmb}
\end{align}
\es
 where $z$ is the coordination number of the bipartite lattice, $J$ is exchange coupling and  $h(t)$ is 
given by Eq.\eqref{shortpulse} with its optimum parameters we used before
for the square lattice case. The subscripts $A$ and $B$ correspond to the
coupled  sublattices. The matrix $\underline{\bm{D}}$ encodes the easy-axis anisotropies by
\be \label{eq:classmatrixD}
\underline{\bm{D}}\, =\begin{pmatrix}
\chi & 0 & 0\\
0 & \chi & 0 \\
0 & 0 & 1
\end{pmatrix}.
\ee
We highlight that the equations \eqref{a:LL} describe only two modes while 
the spin wave description relies on the contribution of all modes in the Brillouin zone.
 
\begin{figure}[htb]
    \centering
    \includegraphics[width=\columnwidth]{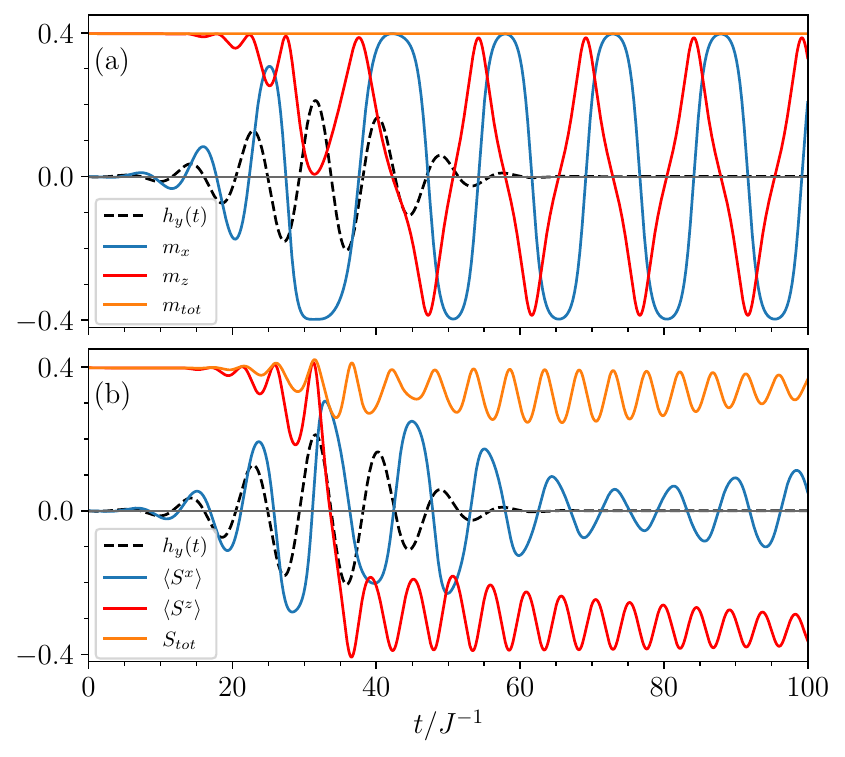}
    \caption{Panel (a): Solid lines represent the dynamics of the classical components of sublattice magnetization including
		the constant total length of the vector since the motion is completely precessional; 
		the black dashed curves shows the pulse. Panel (b):
		Solid lines represent the dynamics of the expectation values of the quantum spin components  including
		their total length $S_{tot}:=\sqrt{\langle S_x\rangle^2+\langle S_y\rangle^2+\langle S_z\rangle^2}$.
		Note the sequence of expectation values and squares; the expectation value of the sum of the squares 
		would yield the constant $S(S+1)$. The black dashed curves show the pulse.
		In both panels, the field amplitude is $h_0=0.214 \,J$; the pulse is applied along the $y$ axis 
		with parameters $\alpha=0.85$, $\phi_0=\pi/3$, and $\tau=10 \, J^{-1}$ in \eqref{shortpulse}; 
		the anisotropy is set to $\chi=0.9$.}
    \label{comp:clasquant}
\end{figure}

In Fig.\ \ref{comp:clasquant}, we compare the classical result obtained from solving
\eqref{a:LL} in the upper panel with the quantum result from spin wave theory in the
lower panel. In the very beginning, the classical and the spin wave result resemble each other, but quickly
strong deviations occur. The most striking feature of the classical simulation is the persistence
of oscillations after the pulse. In contrast, the spin wave result displays significantly weaker
oscillations after the pulse which continue to decrease. This is to be attributed to the dephasing
of all modes in the Brillouin zone. Note that no relaxation, e.g., Gilbert relaxation,
is included in neither of the two calculations.
Another interesting difference consists in the lengths of the vector of the magnetizations.
While the classical magnetizations remain constant in length since they only precess,
the vector of expectation values varies in length, in particular in the vicinity of 
the change of sign in $m_z$. Both differences are clear signatures of the quantum character of the
dynamics.

\end{appendix}	
\end{document}